\documentclass{elsart}
\usepackage{epsfig}
\usepackage{bm}
\usepackage{graphicx}

\newcommand*{\text}{\rm}
\newcommand*{\micron}{\ensuremath{\mu\mathrm{m}}}
\newcommand*{\fb}{\ensuremath{{\text{fb}^{-1}}}}
\newcommand*{\dmd}{\ensuremath{{\Delta m_d}}}
\newcommand*{\bz}{\ensuremath{{B^0}}}
\newcommand*{\bzb}{\ensuremath{{\overline{B}{}^0}}}
\newcommand*{\bzbzbar}{\bz{\rm -}\bzb}
\newcommand*{\pim}{\ensuremath{{\pi^-}}}
\newcommand*{\pip}{\ensuremath{{\pi^+}}}
\newcommand*{\piz}{\ensuremath{{\pi^0}}}
\newcommand*{\ks}{\ensuremath{{K^0_{\rm S}}}}
\newcommand*{\rhom}{\ensuremath{{\rho^-}}}
\newcommand*{\dplus}{\ensuremath{{D^+}}}
\newcommand*{\dstarp}{\ensuremath{{D^{*+}}}}

\newcommand*{\Mevcsq}{\ensuremath{{\rm MeV}/c^2}}
\newcommand*{\Gevc}{\ensuremath{{\rm GeV}/c}}
\newcommand*{\taubz}{\ensuremath{{\tau_{\bz}}}}
\newcommand*{\dt}{\ensuremath{{\Delta t}}}
\newcommand*{\dtp}{\ensuremath{{\dt^\prime}}}
\newcommand*{\zrec}{{z_{\text{rec}}}}
\newcommand*{\zasc}{{z_{\text{tag}}}}
\newcommand*{\dE}{{\Delta E}}
\newcommand*{\mb}{{M_{\rm bc}}}
\newcommand*{\Ebeam}{{E_{\rm beam}^{\rm cms}}}
\newcommand*{\Eb}{{E_B^{\rm cms}}}
\newcommand*{\pb}{{p_B^{\rm cms}}}

\newcommand*{\fsig}{\ensuremath{f_{\rm sig}}}
\newcommand*{\Fsig}{{F_{\rm sig}}}
\newcommand*{\Fbg}{{F_{\rm bkg}}}
\newcommand*{\OF}{{\rm OF}}
\newcommand*{\SF}{{\rm SF}}

\newcommand*{\Pbgof}{\ensuremath{{P^{\OF}_{\rm bkg}}}}

\newcommand*{\Pol}{\ensuremath{{P_{\rm ol}}}}

\newcommand*{\tbg}{\ensuremath{{\tau_{\rm bkg}}}}

\newcommand*{\intl}{29.1}
\newcommand*{\dmdresultmean}{\ensuremath{0.528}}
\newcommand*{\dmdresultstat}{\ensuremath{\pm 0.017}}
\newcommand*{\dmdresultsyst}{\ensuremath{0.011}}
\newcommand*{\dmdresult}{\ensuremath{(\dmdresultmean \dmdresultstat \pm \dmdresultsyst)~\text{ps}^{-1}}}

\begin{document}

\begin{frontmatter}

\hbox to \textwidth{
\lower 2.5cm
\hbox to 2.5cm{
\hss
\epsfysize3cm
\epsfbox{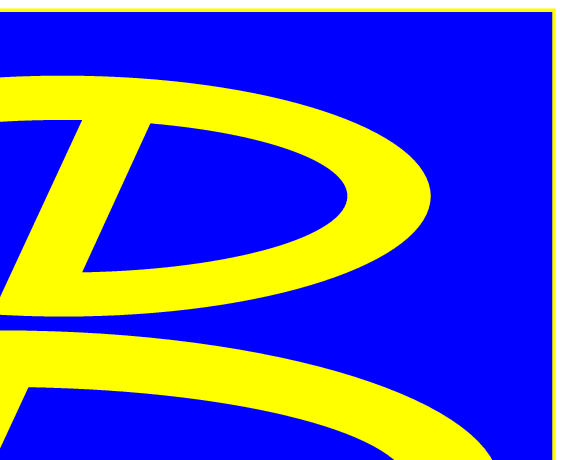}    
}
\hss
\hbox to 3cm{
\begin{tabular}{r}
{KEK preprint 2002-63} \\
{Belle preprint 2002-20}
\end{tabular}
\hss
}
}
\vspace{12pt}

\title{\boldmath Measurement of the Oscillation Frequency for \bzbzbar{} 
  Mixing  using Hadronic \bz{} Decays}

\collab{Belle Collaboration}
  \author[Tokyo]{T.~Tomura}, 
  \author[KEK]{K.~Abe}, 
  \author[TohokuGakuin]{K.~Abe}, 
  \author[TIT]{N.~Abe}, 
  \author[Niigata]{R.~Abe}, 
  \author[Tohoku]{T.~Abe}, 
  \author[KEK]{I.~Adachi}, 
  \author[Tokyo]{H.~Aihara}, 
  \author[Tsukuba]{Y.~Asano}, 
  \author[Toyama]{T.~Aso}, 
% ...

% ...
\address[KEK]{High Energy Accelerator Research Organization (KEK), Tsukuba, Japan}
% ...
\address[Niigata]{Niigata University, Niigata, Japan}
% ...
\address[TohokuGakuin]{Tohoku Gakuin University, Tagajo, Japan}
\address[Tohoku]{Tohoku University, Sendai, Japan}
\address[Tokyo]{University of Tokyo, Tokyo, Japan}
\address[TIT]{Tokyo Institute of Technology, Tokyo, Japan}
% ...
\address[Toyama]{Toyama National College of Maritime Technology, Toyama, Japan}
\address[Tsukuba]{University of Tsukuba, Tsukuba, Japan}
% ...
\thanks[NovaGorica]{on leave from Nova Gorica Polytechnic, Nova Gorica, Slovenia}


\date{\today}

\begin{abstract}
The oscillation frequency of \bzbzbar{} mixing ($\dmd$) has been measured
using \intl{}~\fb{} of data 
collected with the Belle detector at KEKB.
This measurement is made through the distributions of the proper decay 
time difference of $B$ pairs in events tagged as same- and 
opposite-flavor decays.  In each event, one $B$ is fully 
reconstructed in a flavor-specific hadronic decay mode, while the 
flavor of the other is extracted through a likelihood calculated from 
the $b$-flavor information carried in its final decay products.  
We obtain 
$\dmd = \dmdresult$,
where the first error is statistical and the second error is systematic.

\vspace{3\parskip}
\end{abstract}

\end{frontmatter}

\section{Introduction}

In the Standard model,
\bzbzbar{} oscillation occurs through second order weak interactions
where the dominant contribution has an internal loop that contains
virtual $t$-quarks. As a result, the oscillation frequency \dmd{} is related
to the Cabibbo-Kobayashi-Maskawa (CKM) matrix elements
$V_{td}$ and $V_{tb}$~\cite{CKM}. 
Since $|V_{tb}|$ is expected to be very close to 1, 
a precise measurement of \dmd{}, in principle,  provides a method 
to determine $|V_{td}|$. 
 Although
  the extraction of $|V_{td}|$ from \dmd{} is currently hampered by 
  theoretical uncertainties on hadronic matrix elements,
precision measurements of \dmd{} could be important for future
determinations of $|V_{td}|$.
In addition, \bzbzbar{} mixing is one of the sources of $CP$ violation
in \bz{} decays 
and a good knowledge of \dmd{} is important 
for precise measurements of these $CP$ asymmetries.
Since the first observation of \bzbzbar{} mixing \cite{Argus_mix}, 
a number of measurements of \dmd{} have been reported~\cite{dmd_wg}.  
Recently, asymmetric $B$-factory experiments operated at the $\Upsilon(4S)$
resonance have 
significantly improved the precision
of \dmd{}~\cite{ABF_dmd}.

In this letter,  we present a determination of \dmd{} from the time evolution
of opposite-flavor (OF; \bz{}\bzb{}) and same-flavor 
(SF; \bz{}\bz{}, \bzb{}\bzb{})
neutral $B$ decays at the $\Upsilon(4S)$ resonance.
In the absence of background and detector effects,
the time-dependent probabilities of observing 
OF ($\mathcal{P}^{\OF}$)  
and SF ($\mathcal{P}^{\SF}$) 
states are given by
\begin{eqnarray}
  \label{eqn:true_unm}
  \mathcal{P}^{\OF}(\dt) &=& 
   \frac{1}{4\taubz} \exp\left( -\frac{|\dt|}{\taubz} \right)
   \left[ 1 + \cos(\dmd\dt) \right] ,\\
  \label{eqn:true_mix}
  \mathcal{P}^{\SF}(\dt) &=& 
   \frac{1}{4\taubz} \exp\left( -\frac{|\dt|}{\taubz} \right)
   \left[ 1 - \cos(\dmd\dt) \right] ,
\end{eqnarray}
where $\taubz$ is the $\bz$ lifetime,
and $\dt$ is the proper time difference between the two $B$ meson decays.
As the decay width difference of the two mass eigenstates of
the $\bz\bzb$ system is expected to be very small 
in the Standard Model~\cite{DG_SM},
we assume in this analysis that it is equal to zero.

The analysis described here is based on a \intl{}~\fb{} data sample,
which contains $31.3 \times 10^6$ $B\overline{B}$ pairs, 
collected with the Belle
detector~\cite{Belle} at the asymmetric-energy KEKB storage ring~\cite{KEKB}.
KEKB collides an 8.0~GeV electron beam and a 3.5~GeV positron beam
with a crossing angle of 22 mrad,
resulting in a center-of-mass system (cms) moving nearly along the 
$z$ axis (defined as anti-parallel to the positron beam)
with a Lorentz boost of $(\beta\gamma)_\Upsilon = 0.425$.
Since $B$ mesons are nearly at rest in the $\Upsilon(4S)$ cms,
the proper time difference $\dt$ can be approximated as
$  \dt \simeq \Delta z / [c (\beta \gamma)_\Upsilon], $
where $\Delta z$ is the distance between the decay vertices
of the two $B$ mesons in $z$ (typically $200~\micron$).
The measurement involves the reconstruction of the decay of
a neutral $B$ meson in a flavor-specific hadronic mode,
the determination of the $b$-flavor of the accompanying (tagging)
$B$ meson, the reconstruction of the two decay vertices to determine $\dt$,
and a fit of the $\dt$ distribution taking into account
$\dt$ resolution and backgrounds to extract $\dmd$.  
We use the same flavor-tagging method as for the $\sin2\phi_1$ 
measurement \cite{CP1_Belle}, and the same vertex reconstruction
and resolution parameters as those used in the lifetime 
measurements \cite{Belle_blife}.

The Belle detector consists of
a three-layer silicon vertex detector~(SVD),
a 50-layer central drift chamber~(CDC),
an array of  
aerogel Cherenkov counters~(ACC),  
time-of-flight scintillation counters~(TOF),
an electromagnetic calorimeter containing  
CsI(Tl) crystals~(ECL),
and 14 layers of 4.7 cm thick iron plates interleaved with
a system of resistive plate counters~(KLM).
All subdetectors except the KLM are located inside
a 3.4~m diameter superconducting solenoid which provides
a 1.5~T magnetic field.
The impact parameter resolutions for charged tracks are measured to be
$\sigma_{xy}^2 = (19)^2 + (50/(p\beta\sin^{3/2}\theta))^2~\micron^2$
in the plane perpendicular to the $z$ axis and
$\sigma_{z}^2 = (36)^2 + (42/(p\beta\sin^{5/2}\theta))^2~\micron^2$
along the $z$ axis, where $\beta = pc/E$, $p$ and $E$ are
the momentum ($\Gevc$) and energy (GeV) of the particle, and
$\theta$ is the polar angle from the $z$ axis.

\section{Event Selection and Reconstruction}

$\bzb$ mesons are fully reconstructed in the decay modes\footnote{
 The inclusion of the charge conjugate decays is implied
 throughout this letter. }
$\bzb \to \dplus\pim$, $\dstarp\pim$, and $\dstarp\rhom$.  
Charged pion and kaon candidates are required to satisfy
selection criteria based on particle-identification likelihood functions
derived from specific ionization (d$E$/d$x$) in the CDC,
time of flight,  
and the response of the ACC.
Photon candidates are defined as isolated ECL clusters with energy more than
30~MeV that are not matched to any charged track.  
We reconstruct $\piz$ candidates
from pairs of photon candidates
with invariant masses between 124 and $146~\Mevcsq$.
A mass-constrained fit is performed to improve the $\piz$ momentum resolution.
A minimum $\piz$ momentum of $0.2~\Gevc$ is required.
We select $\rhom$ candidates
as $\pim\piz$ pairs having invariant masses
within $\pm 150~\Mevcsq$ of the nominal $\rhom$ mass.

Neutral and charged $D$ candidates are reconstructed
in the following channels:
$D^0 \to K^-\pip$, $K^-\pip\piz$, $K^-\pip\pip\pim$, and $D^+ \to K^-\pip\pip$.
Candidate $\dstarp \to D^0\pip$ decays are formed by combining
a $D^0$ candidate with a slow and positively charged track,
for which no particle identification is required.
We apply mode-dependent requirements on the reconstructed $D$ mass 
(ranging from $\pm 15$ to $\pm 58$ MeV/$c^2$) and the mass difference
between $D^{*+}$ and $D^0$ (ranging from $\pm 3$ to $\pm 12$ MeV/$c^2$).
To reduce continuum background, a mode-dependent selection is applied
based on the ratio of the second to zeroth Fox-Wolfram moments~\cite{FW}
and the angle between the thrust axes of the reconstructed and
associated $B$ mesons.

We identify $B$ decays based on requirements on
the energy difference $\dE \equiv \Eb - \Ebeam$ and
the beam-energy constrained mass $\mb \equiv \sqrt{(\Ebeam)^2-(\pb)^2}$,
where $\Eb$ and $\pb$ are the cms energy
and momentum of the fully reconstructed $B$ candidate, and
$\Ebeam$ is the cms beam energy.
If more than one fully reconstructed $B$ candidate is found
in the same event, the one with the best combined $\chi^2$ for $\dE$, $\mb$,
and the invariant mass of the $D$ candidate is chosen.
For each channel, a rectangular signal region is defined in the $\dE$-$\mb$
plane, corresponding to $\pm 3\sigma$ windows centered on the expected
means for $\dE$ and $\mb$.
The $\dE$ resolution is $\sigma = 10\sim 30$~MeV, 
depending on the decay mode and
$\sigma\simeq 3~\Mevcsq$ for $\mb$ for all modes (5.27 $< \mb <$ 5.29 GeV/$c^2$
is used).
For the determination of background parameters,
we use candidates in a background dominated (sideband) region,
$-0.1$ ($-0.05$ for $D^{*+}\rho^-$) $< \dE <$ 0.2 GeV 
and 5.20 $< \mb <$ 5.29 GeV/$c^2$, and exclude the signal region.

Charged leptons, pions, and kaons
that are not associated with the reconstructed
hadronic decay are used to identify
the flavor of the accompanying $B$ meson.
We apply the same method that has been used for our
measurement of $\sin 2\phi_1$~\cite{CP1_Belle}.
Initially,
the $b$-flavor determination is performed at the track level.
Several categories of well measured tracks
that 
have a charge correlated with the $b$ flavor are selected:
high momentum leptons
from  $b\to$ $c\ell^-\overline{\nu}$,
lower momentum leptons from  $c\to$ $s\ell^+\nu$,
charged kaons and $\Lambda$ baryons from $b\to$ $c\to$ $s$,
high momentum pions originating from decays of the type
$B^0\to$ $D^{(*)-}X$ (where $X = \pi^+, \rho^+$, $a_1^+, {\rm etc.})$, and
slow pions from $D^{*-}\to$ $\overline{D}{}^0\pi^-$.
We use Monte Carlo (MC) simulations to determine a category-dependent 
variable
that indicates whether
a track originates from a $B^0$ or $\overline{B}{}^0$.
The values of this variable range from 
$-1$ for a reliably identified $\overline{B}{}^0$
to $+1$ for a reliably identified $B^0$ and  depend
on the tagging particle's  charge,
cms momentum, polar angle and
particle-identification probability, as well as other kinematic and
event shape quantities. 
The results from the separate
track categories are then combined to 
take into account correlations
in the case of multiple track-level tags.
We use two parameters, $q$ and $r$, to represent the tagging information.
The first, $q$, corresponds to the sign of
the $b$ quark charge of the tag-side $B$ meson 
where $q = +1$ for $\overline b$ and hence
$B^0$, and $q = -1$ for $b$ and $\bzb$.
The parameter $r$ is an event-by-event,
MC-determined flavor-tagging dilution factor 
that ranges from $r=0$ for no flavor
discrimination to $r=1$ for unambiguous flavor assignment.
It is used only to sort data into six intervals of $r$
(boundaries at 0.25, 0.5, 0.625, 0.75, 0.875), according to 
flavor purity. 
More than 99.5\% of the events are assigned a non-zero value of $r$.

The decay vertices of the two $B$ mesons in each event are fitted
using tracks that have at least one SVD hit
in the $r$-$\phi$ plane and at least two SVD hits in the $r$-$z$ plane,
under the constraint that they are consistent with
the interaction point (IP) profile, smeared in the $r$-$\phi$ plane
by $21~\micron$ to account for the transverse $B$ decay length.
The IP profile is represented by a three-dimensional Gaussian,
the parameters of which are determined in each run
(every 60,000 events in case of the mean position)
using hadronic events.
The size of the IP region is typically $\sigma_x \simeq 100~\micron$,
$\sigma_y \simeq 5~\micron$, and $\sigma_z \simeq 3$~mm,
where $x$ and $y$ denote horizontal and vertical directions, respectively.

The decay point of the reconstructed $\bz$ is obtained from the vertex 
position and momentum vector of the reconstructed $D$ meson and
a track (other than the slow $\pip$ candidate from $\dstarp$ decay)
that originates from the fully reconstructed $B$ decay.
The decay vertex of the tagging $B$ meson is determined
from tracks not assigned to the fully reconstructed $B$ meson;
however, poorly reconstructed tracks (with a longitudinal position error
in excess of $500~\micron$) as well as tracks likely to come from
$\ks$ decays 
(because they either form a $\ks$ mass when combined with an another 
oppositely charged track
or miss the fully reconstructed $B$ vertex by 
more than $500~\micron$ in the $r$-$\phi$ plane) are not used.

The quality of a fitted vertex is assessed only in the $z$ direction
(because of the tight IP constraint in the transverse plane),
using the variable  
$\xi \equiv (1/2n) \sum^n_i \left[ (z_{\rm after}^i-z_{\rm before}^i) /
  \varepsilon_{\rm before}^i \right]^2$,
where $n$ is the number of tracks used in the fit,
$z_{\rm before}^i$ and $z_{\rm after}^i$ are the $z$ positions
of each track (at the closest approach to the origin)
before and after the vertex fit, respectively,
and $\varepsilon_{\rm before}^i$ is the error of $z_{\rm before}^i$.
A MC study shows that $\xi$ does not depend
on the $B$ decay length.
We require $\xi < 100$ to eliminate poorly reconstructed vertices.
About 3\% of the fully reconstructed vertices
and 1\% of the tagging $B$ decay vertices are rejected.
The proper-time difference between the fully reconstructed
and the associated $B$ decays, $\dt \equiv t_{\rm rec} - t_{\rm tag}$, 
is calculated
as $\dt =(\zrec - \zasc)/[c(\beta\gamma)_\Upsilon]$, where $\zrec$ and $\zasc$
are the $z$ coordinates of the fully-reconstructed and associated
$B$ decay vertices, respectively.
We reject a small fraction ($\sim$0.2\%) of the events by requiring
$|\dt| < 70$~ps ($\sim$45$\taubz$).

\begin{figure}
  \resizebox{0.6\textwidth}{!}{\includegraphics{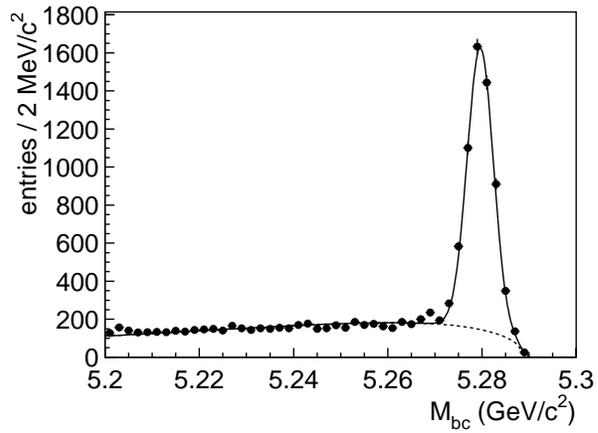}}
  \caption{\label{fig:mbc}
    Beam-energy constrained mass distribution of the fully
    reconstructed $B^0$ candidates.
    The dashed curve shows the background contributions
    and the solid curve shows the sum of signal and background.}
\end{figure}

 Figure~\ref{fig:mbc} shows the $\mb$ distribution for all
the candidates found in the $\dE$ signal region
after flavor tagging and vertex reconstruction.  
We find 
 2269 $D^+\pi^-$, 2490 $D^{*+}\pi^-$, and 1901 $D^{*+}\rho^-$ 
candidates in the signal region
 with average purities of 86\%, 81\%, and 70\%, respectively.
The flavor-tagging intervals $l = 1, 2, \cdots 6$ contain
2675, 981, 597, 702, 779, and 926 candidates, respectively.

\section{\label{sec:mixingfit} 
   {\boldmath Extraction of $\Delta m_{\bf\it d}$}}

An unbinned maximum likelihood 
fit is performed to extract \dmd{},
based on a likelihood function defined as
$ L(\dmd, w_1, w_2, \cdots, w_6)  
  = \prod_i \tilde P^{\OF}(\dt_i) \times 
                 \prod_j \tilde P^{\SF}(\dt_j)$,
where the index $i$ ($j$) runs over all selected OF(SF) events in the 
signal region.
The probability density function 
(PDF)\footnote{Note that the PDF is normalized as 
 $ \int [ \tilde P^{\OF}(\dt) + \tilde P^{\SF}(\dt) ] \mathrm{d}(\dt) = 1. $ },
$\tilde P^{\OF(\SF)}$, 
is expressed as
\begin{eqnarray}
  \tilde P^{\OF(\SF)}(\dt) &=& 
   \fsig [ (1-f_{\rm ol}) P_{\rm sig}^{\OF(\SF)}(\dt; \dmd, w_l) 
      + f_{\rm ol} f_{\rm sig}^{\OF(\SF)} P_{\rm ol}(\dt) ]
 \nonumber \\
   &+& (1-\fsig ) [ (1-f_{\rm ol}) P_{\rm bkg}^{\OF(\SF)}(\dt)
      + f_{\rm ol} f_{\rm bkg}^{\OF(\SF)} P_{\rm ol}(\dt) ],
\end{eqnarray}
where 
$\fsig$ is a signal purity, described below, and 
$w_l$ is the wrong tag fraction 
of the flavor-tagging interval $l$ containing event $i$ or $j$.
The signal PDF 
($P^{\OF(\SF)}_{\rm sig}$)  
is 
the convolution of the true PDF with the resolution function ($R_{\rm sig}$).
The background PDF ($\Pbgof$$^{(\SF)}$) 
is expressed in a similar way:
\begin{equation}
  P_k^{\OF(\SF)}(\dt) = \int_{-\infty}^{+\infty} \mathrm{d}(\dtp)
  \mathcal{P}_k^{\OF(\SF)}(\dtp) R_k(\dt - \dtp) ,
\end{equation}
where $k$ = sig, bkg.  
A small number of signal and background events have large $\dt$ 
values.  We account for the contribution from these ``outliers'' 
by adding a Gaussian component $\Pol(\dt)$ with a width and contribution 
fraction determined from our $B$ lifetime analysis~\cite{Belle_blife}.
The width of the outlier Gaussian is
$\sigma_{\rm ol}$ = 36 ps,
and the fraction, $f_{\rm ol}$, is 0.06\% if both 
  vertices are reconstructed from two or more 
  charged tracks and
  3.1\% if at least one vertex is reconstructed from a 
  single track constrained with the IP profile.
We assume $f_{\rm ol}$ to be the same for signal and background 
  as well as for the OF and SF sub-samples.
$f_k^{\OF(\SF)}$ is the fraction of OF(SF) events for $k$ (= sig, bkg)
and given by $N_k^{\OF(\SF)}/(N_k^{\OF}+N_k^{\SF})$,
where $N_k^{\OF(\SF)}$ is the number of OF(SF) events in the signal
region for $k$
from the result of the $\dE{}$-$\mb{}$ distribution fit described below.

The signal purity ($\fsig$) 
is determined on an event-by-event basis
as a function of $\dE$ and $\mb$.
The two-dimensional distribution of these variables including
the sideband region is fitted with the sum of
a Gaussian function $\Fsig(\dE,\mb)$ to represent the signal, and
a background function $\Fbg(\dE,\mb)$, represented as
an ARGUS background function~\cite{ARGUS_bkg} in $\mb$ and
a first-order polynomial in $\dE$.
The fraction of signal is taken to be
$\fsig(\dE,\mb) = \Fsig/[\Fsig+\Fbg]$.
The fits are done separately for the three $B$ decay modes, and
the dependence of the signal fraction on the flavor-tag interval $l$ is
included in the overall normalization of $\fsig$. 

In order to account for wrong tagging,
the true signal PDF 
($\mathcal{P}_{\rm sig}^{\OF(\SF)}$)
is given by replacing
$\cos(\dmd\dt)$ with $(1-2w_l)\cos(\dmd\dt)$ in 
Eq.~\ref{eqn:true_unm} (Eq.~\ref{eqn:true_mix})\footnote{
 We neglect the $\sin(\dmd\dt)$ term that may arise due to the 
 interference between Cabibbo-favored and suppressed decay 
 amplitudes~\cite{Dunietz},
 since its effect is expected to be very small. }.
%
The resolution function of the signal is constructed by
convolving four different contributions:
the detector resolutions on $\zrec$ and $\zasc$,
the smearing of $\zasc$ due to the inclusion of tracks
which do not originate from the associated $B$ vertex, 
mostly due to charm and $K_{\rm S}$ decays,
and the kinematic approximation that the $B$ mesons are at rest
in the cms. 
The resolution is determined on an event-by-event basis, using the
estimated uncertainties on the vertex $z$ positions determined 
from the vertex fit.  
The parameterization of the resolution depends on whether the vertices 
are reconstructed with multiple tracks or a single track.
A detailed description of the resolution parameterization can be found in
Ref.~\cite{Belle_blife}.
We use the same parameters for $R_{\rm sig}(\dt)$ 
obtained there for this analysis. 
The average $\dt$ resolution for the signal is $\sim$1.56~ps (rms).

The background PDF is modeled as a sum of exponential
and prompt components,
\begin{eqnarray}
 \mathcal{P}_{\rm bkg}^{\OF(\SF)}(\dt) &=& 
  f^{\OF(\SF)}_{\rm bkg} \left[ 
  f^\delta_{\rm bkg}\delta(\dt-\mu_\delta^{\rm bkg}) +
  \frac{(1- f^\delta_{\rm bkg})}{ 2\tbg }
    e^{-|\dt-\mu_\tau^{\rm bkg}|/{\tbg} }
  \right] , 
\end{eqnarray}
 convolved with 
$R_{\rm bkg}(\dt)$, which is parameterized as a sum of two Gaussians.
Different parameter values are used for $R_{\rm bkg}(\dt)$
depending on whether or not both vertices are reconstructed
with multiple tracks.
The parameters for the background PDF  
are determined using
the $\dE$-$\mb$ sideband region for each decay mode.
A MC study shows that the fraction of prompt component
in the signal region is smaller (by $\sim$10--50\%
depending on the decay mode) than that in the sideband region.
We correct the estimates of $f_{\rm bkg}^{\delta}$
for this effect.

\begin{figure}
  \resizebox{0.60\textwidth}{!}{\includegraphics{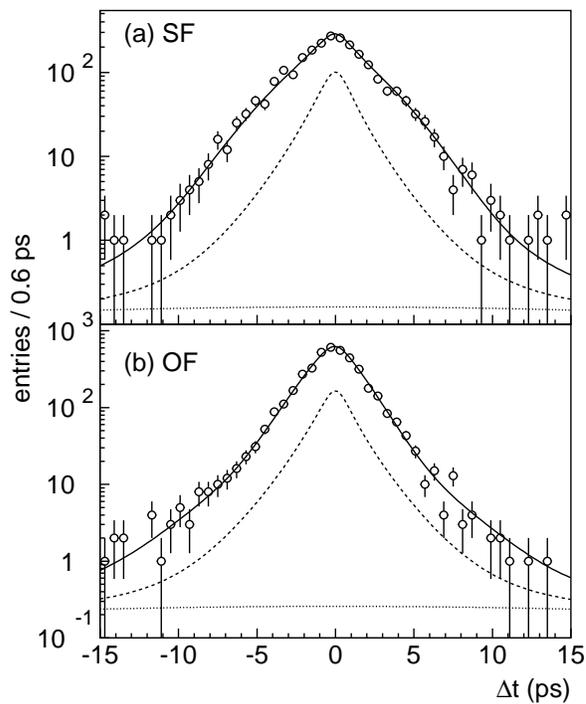}}
  \caption{\label{fig:fit_result}
   Distributions of $\dt$ for (a) SF and (b) OF events with the fitted 
   curves superimposed.  The dashed, dotted, and solid curves show 
   the background, outliers, and the sum of backgrounds and signal, 
   respectively.
  }
\end{figure}

\begin{figure}
  \resizebox{0.6\textwidth}{!}{\includegraphics{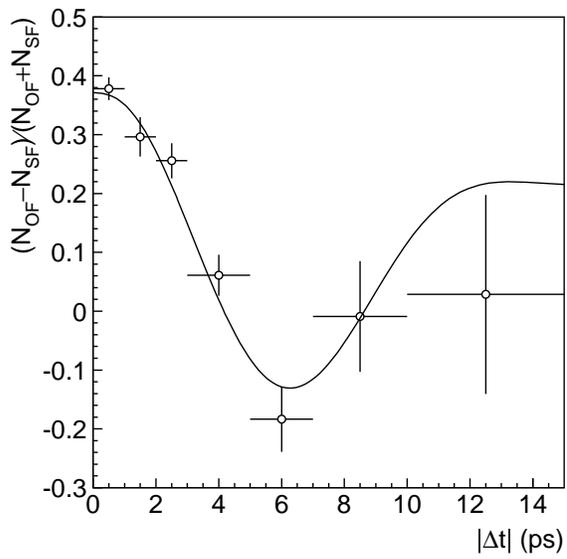}}
  \caption{\label{fig:asym}
    Time dependence of the asymmetry between
    OF and SF events.  The curve shows the result of the $\dmd$ fit.
   }
\end{figure}

In the final fit, we fix \taubz{} to the world average value \cite{PDG}
and determine \dmd{} and $w_l$ ($l$=1,6).  The fit result is
listed in Table~\ref{tab:fit_result}.  Separate fits to the
$\dplus\pim$, $\dstarp\pim$, and $\dstarp\rhom$ decay modes give
consistent $\dmd$ values:
0.536 $\pm$ 0.027 ps$^{-1}$, 0.543 $\pm$ 0.027 ps$^{-1}$, and
0.497 $\pm$ 0.032 ps$^{-1}$, respectively.
Figure~\ref{fig:fit_result} shows the $\dt$ distributions 
for OF and SF events with the fitted curves superimposed;
Fig.~\ref{fig:asym} shows the asymmetry between OF and SF events,
$(N_{\OF} - N_{\SF})/(N_{\OF} + N_{\SF})$,
as a function of $|\dt|$.

The systematic errors are summarized in Table~\ref{tab:syserror}.
The dominant sources are the uncertainties in the resolution functions.
The fit is repeated after varying the parameters
determined from the data (MC) by $\pm 1\sigma$ ($\pm 2\sigma$).
The systematic error due to the modeling of the resolution function 
is estimated by comparing the results with different 
parameterizations.
The systematic error due to the IP constraint is estimated
by varying ($\pm 10~\micron$) the smearing used to account for
the transverse $B$ decay length.
The IP profile is determined using two different methods and
we find no difference between the results.
Possible systematic effects due to the track quality selection
of the associated $B$ decay vertices are studied by varying
each criterion by 10\%.
The fit quality criterion for reconstructed vertices is varied
from $\xi < 50$ to $\xi < 200$.
We check the systematic uncertainty due to 
outliers  
by varying the  
$\dt$ range to $\pm 40$~ps and $\pm 100$~ps, and find
a negligibly small effect.
$\dE$-$\mb$ signal regions are varied by $\pm 10$~MeV for $\dE$
and $\pm 3~\Mevcsq$ for $\mb$.
The parameters determining $\fsig$ are varied by $\pm 1\sigma$
to estimate the associated systematic error.
We study background components with a MC sample that includes both
$B\overline{B}$ and continuum events.  We find no significant peaking
background in the signal region above the fitted background
curve and conclude that the effect of peaking background
is negligibly small. 
The systematic error due to the background shape is estimated
by varying its parameters by their errors.
In the nominal fit, we do not include any oscillation component in
the background; such a component  may arise from 
the $\bz$ originated background.
We repeat the fit with a background PDF including 
a mixing term,
where background parameters are taken from 
the sideband data.
The dependence on the $\bz$ lifetime
is measured by varying the lifetime by $\pm 1\sigma$ from the world average
value.
The possible bias in the fitting procedure 
and the effect of SVD alignment error are studied with MC samples;
we find no bias.  
The MC statistical error is associated as a systematic error 
for these sources.

\section{ Conclusion }
We have presented a new measurement of $\dmd$ 
using $\intl~\fb$ of data collected with the Belle detector
at the $\Upsilon(4S)$ energy.
An unbinned maximum likelihood fit to the distribution of the proper-time
difference of a flavor-tagged sample with one of the neutral $B$ mesons 
fully reconstructed in hadronic decays yields
$$ \dmd = \dmdresultmean \dmdresultstat {\rm (stat)}
          \pm \dmdresultsyst {\rm (syst)} {\rm ~~ps}^{-1}. $$
This result has similar precision and is consistent with other recent 
measurements at asymmetric $B$-factories 
at the $\Upsilon(4S)$~\cite{ABF_dmd}, which are achieving  
higher precision than those at higher energies. 
Additionally, this measurement confirms the validity of the
$\sin2\phi_1$ measurement performed on the same data sample~\cite{CP1_Belle},
since it is based on the same flavor-tagging method,
and vertexing and fitting procedures.

\section*{Acknowledgments}

We wish to thank the KEKB accelerator group.
We acknowledge support from the Ministry of Education,
Culture, Sports, Science, and Technology of Japan
and the Japan Society for the Promotion of Science;
the Australian Research Council
and the Australian Department of Industry, Science and Resources;
the National Science Foundation of China under contract No.~10175071;
the Department of Science and Technology of India;
the BK21 program of the Ministry of Education of Korea
and the CHEP SRC program of the Korea Science and Engineering Foundation;
the Polish State Committee for Scientific Research
under contract No.~2P03B 17017;
the Ministry of Science and Technology of the Russian Federation;
the Ministry of Education, Science and Sport of Slovenia;
the National Science Council and the Ministry of Education of Taiwan;
and the U.S.\ Department of Energy.


\newpage
\begin{table}[h]
  \caption{\label{tab:fit_result} Summary of fit result.}
    \begin{tabular}{cccc}
      \hline
     Fit parameter & Fit value & Fit parameter & Fit value \\
     \hline
     $\dmd$ & $\dmdresultmean\dmdresultstat$ ps$^{-1}$ & & \\
     $w_1$ & 0.478 $\pm$ 0.017 & $w_2$ & 0.313 $\pm$ 0.027 \\
     $w_3$ & 0.212 $\pm$ 0.030 & $w_4$ & 0.187 $\pm$ 0.027 \\
     $w_5$ & 0.088 $\pm$ 0.022 & $w_6$ & 0.016 $\pm$ 0.013 \\
     \hline
    \end{tabular}
\end{table}

\vspace{1cm}
\begin{table}[h]
  \caption{\label{tab:syserror} Summary of systematic errors.}
    \begin{tabular}{lc}
      \hline
      Source                   & Error (ps$^{-1}$) \\
      \hline
      Resolution parameters    & $0.008$           \\
      Resolution parameterizations & $0.003$           \\
      IP constraint            & $0.001$           \\
      Track selection          & $0.002$           \\
      Vertex selection         & $0.003$           \\
      $\dE$-$\mb$ signal box   & $0.001$           \\
      Signal fraction          & $0.001$           \\
      Background shape         & $0.002$           \\
      Mixing in the background & $0.002$           \\
      $\bz$ lifetime           & $0.002$           \\
      Fit bias                 & $0.005$           \\
      \hline
      Total                    & \dmdresultsyst{}  \\
      \hline
    \end{tabular}
\end{table}

\end{document}